# The effect of rippling on the mechanical properties of graphene

*Guillermo Lopez-Polin[1,2]\*, Cristina Gomez-Navarro[2,3], Julio Gomez-Herrero[2,3]*

[1] Instituto de Ciencia de Materiales de Madrid, CSIC, Madrid E 28049, Spain
[2] Departamento de Física de la Materia Condensada, Universidad Autónoma de Madrid, Madrid E-28049, Spain
[3] IFIMAC Condensed Matter Physics Center (IFIMAC). Universidad Autónoma de Madrid, Madrid E-28049, Spain

\*E-mail: guillermo.lp@csic.es

**Abstract**

Graphene is the stiffest material known so far but, due to its one-atom thickness, it is also very bendable. Consequently, free-standing graphene exhibit ripples that has major effects on its elastic properties. Here we will summarize three experiments where the influence of rippling is essential to address the results. Firstly, we observed that atomic vacancies lessen the negative thermal expansion coefficient of free-standing graphene. We also observed an increase of the Young's modulus with global applied strain and with the introduction of small density defects that we attributed to the decrease of rippling. Here, we will focus on a surprising feature observed in the data: the experiments consistently indicate that only the rippling with wavelengths between 5-10nm influences the mechanics of graphene. The rippling responsible of the negative TEC and anomalous elasticity is thought to be dynamic, i.e. flexural phonons. However, flexural phonons with these wavelengths should have minor effects on the mechanics of graphene, therefore other mechanisms must be considered to address our observations. We propose static ripples as one of the key elements to correctly understand the thermomechanics of graphene and suggest that rippling arises naturally due to a competition of symmetry breaking and anharmonic fluctuations.

**Introduction**

In mechanics, membranes are defined as elements with pure in-plane elastic response. Accordingly, two dimensional materials, and specifically graphene, are the paradigm of membranes. Graphene, a one-atom thick layer of carbon atoms, exhibits the highest known in-plane stiffness and an extremely low bending rigidity [1, 2]. As a consequence, thermal energy at room temperature is enough to induce significant out of plane thermal fluctuations on graphene membranes. In fact, Mermin theorem predicted that these flexural phonons were strong enough to destroy the crystalline order of a graphene monolayer [3]. However, the anharmonic (nonlinear) coupling between bending and stretching modes attenuate out of plane phonons making possible the existence of graphene [4]. These thermally activated flexural phonons are assumed to have a strong influence on the mechanical and thermomechanical properties of membranes [5, 6].

The first experimental observation of materials where thermal fluctuations played a dominant role in defining the elastic constants were biological membranes such as lipid bilayers [7]. These membranes show renormalization of the elastic constants and thermal contraction due to flexural phonons [8]. To get an intuitive idea on the influence of flexural phonons in a membrane we can make an analogy of a fluctuating membrane with a sheet of paper with static wrinkles as exemplified in Figure 1; stretching a wrinkled paper requires less energy than when it is flat (figure 1.a and c). Also, if you hold firmly one corner of a flat paper sheet it will bend by its own weight, contrary to a wrinkled paper sheet that stands alone (Figure 1.b and d). Finally, if you wrinkle a paper sheet, the projection of the sheet on a plane would be smaller. Likewise, out of plane fluctuations reduce the in-plane stiffness, increase the bending rigidity and cause negative thermal coefficient to these membranes. Dynamic fluctuations and static wrinkling have similar effects. Consequently, in some cases, their effects are difficult to differentiate. From now on we will use the term rippling (or ripples) to denote both effects.

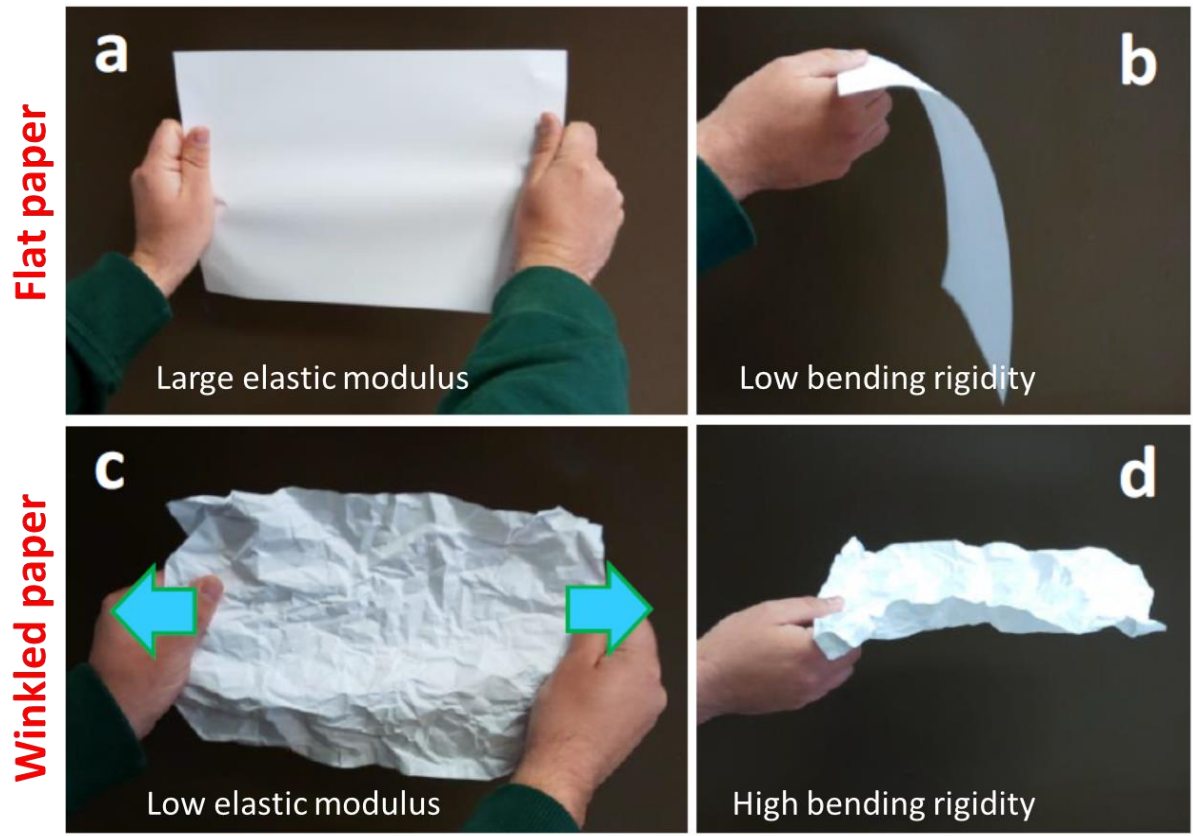

*Figure 1. Illustration of how rippling affects different elastic properties of membranes. A sheet of paper presents different elastic constant depending on its wrinkled or plane structure. a) When flat it presents large resistance to stretching i.e. large elastic modulus. b) If you hold it from one side it will bend easily. i.e low bending rigidity. c) If wrinkled, the same sheet of paper will be much easier to stretch than when it is plane i.e. low elastic modulus. d) When holding the wrinkled sheet from one side it will not bend (increased bending rigidity).*

Graphene and other atom-thick materials have a bending rigidity around 1eV, comparable to that of biological membranes. Hence, a proper understanding of the mechanical properties of these materials should include not only the structural rigidity and configuration of the covalent bonds, but also the effect of rippling. There are several facts that demonstrate that rippling exists in 2D materials and have significant consequences on its elastic response. The most accepted effect is the negative thermal expansion coefficient (TEC) of graphene and h-BN (hexagonal Boron Nitride) that has been observed experimentally [9, 10] and studied in theoretical works [11-13].

The influence of rippling on the elasticity of these materials is still a topic under research. It is well-accepted that micron scaled wrinkles that appear spontaneously during transfer and preparation processes reduce the Young´s modulus and increase the bending rigidity of graphene [14-16]. In contrast to these micron scale wrinkles, the effect of nanometer scaled ripples, as measured by TEM in suspended graphene [17], is highly controversial. The physics behind these ripples remains unclear and debated. So, one can find different theories about their origin [18-23].

Here we present an overview of three experiments carried out by us where the influence of rippling is essential to understand the results [24-26]. In the first set of experiments, we studied the variation of the TEC of graphene (the fingerprint of rippling) with the density of induced single atom vacancies. We observed a decrease of the thermal contraction of graphene with the density of defects. Molecular dynamic simulations indicated that the local strain created around the single vacancies was high enough to quench rippling. In the second set of experiments, we measured the Young´s modulus of graphene as a function of the density of induced atomic vacancies. We observed an initial increase of the Young´s modulus with the density of vacancies. This effect was attributed to the reduction of ripples in graphene. By last, we determined the dependence of the Young´s modulus with pressure-induced global strain on graphene blisters. We observed an increase of the Young's modulus of graphene with increasing strain in the membrane, until a certain tension where it tends to a figure that is twice the initial one. As strain should also remove ripples, we also attributed this stiffening to rippling decrease. To finish up, we will compare the three experimental results and discuss that these effects cannot be exclusively explained in terms of thermal fluctuations. In these works we pointed towards dynamic fluctuations as the main mechanism for the observations although we did not discard static rippling. However, recent theoretical works [20, 27], experiments [28] and some details of our data supports intrinsic static rippling as the governing mechanism for the renormalization of the elastic constants and the negative TEC of graphene. These experiments were already published in references [24-26]. In this work, we intend to shed light on the same data but with a complementary viewpoint.

**Methods**

There is a common methodology used in the experiments commented here based on AFM nanoindentations and defect induction by $Ar^+$ irradion that we describe briefly in the forthcoming lines.

The samples were prepared by mechanical exfoliation of graphite on $SiO_2$(300nm)/Si substrates with predefined circular cavities. For the study, we selected only single layer membranes as confirmed by optical contrast and Raman spectroscopy [29]. The graphene flakes covered these holes resulting in circular graphene drumheads. Using the tip of an Atomic Force Microscope (AFM) we performed indentations on the center of the circular membranes and obtained force ($F$) vs. indentation ($\delta$) curves. These curves presented a cubic behavior described by the equation [1, 30, 31]:

$$F(\delta) = \frac{2\pi\sigma_0}{\ln(a/R_{tip})}\delta + q(v)\frac{E_{2D}}{a^2}\delta^3 \quad (1)$$

Where $\sigma_0$ is the prestress of the membrane, $E_{2D}$ is the 2D Youngs modulus, $a$ is the radius of the circular rim, $R_{tip}$ is the tip radius and $q(v)$ is a factor that depends only on the Poissons coefficient ($v$) of the membrane. The factor $q(v)$ was numerically obtained to make the equation valid for every $v$, and has a value of $q(v) = 1/(1.049 - 0.15v - 0.16v^2)$ [30]. To obtain the results presented here, we used the theoretical value $v = 0.165$ for graphene [32], which yields $q(v) \sim 1$.

Indentation is not a direct experimental measure; it is calculated from the difference between the total displacement on the membrane and the cantilever deflection on the non-deforming $SiO_2$ substrate. As the cantilever is bended during the experiments, the result of this difference is the net displacement of the membrane (the membrane indentation). The correct determination of the contact point between the AFM tip and the membrane is crucial for this analysis as it defines the zero value for $\delta$. Inaccuracy of 2-5 nm in this point leads to a 10% error in $E_{2D}$. As proposed previously [33], fitting the indentation curves to a full cubic polynomial drastically reduce the error as a function of the inaccuracy on the determination of the contact point. Therefore, for the determination of the Young´s modulus we always fitted the curves to a full third order polynomial and used the cubic coefficient:

$$F = C_0 + C_1\delta + C_2\delta^2 + C_3\delta^3 \qquad (2)$$

A more detailed description of curve fitting can be found in reference [24] and [26].

We induced a controlled type and density of defects on the graphene membranes by low energy (140eV) Argon ion bombardment [34, 35]. We estimated in situ the density of defects created by measuring the total charge accumulated in the sample during ion irradition and assuming that every ion impact removes one carbon atom from the graphene lattice [34-36]. Later, we determined the density of defects by Raman spectroscopy, confirming an efficiency of almost 100%. From Raman spectra we can also obtain information about the nature of defects. To this end, we monitored the ratio between D´ and D peaks for different irradiation doses, which was approximately 7, indicating $sp^2$ defects [37]. To obtain further information on the nature of the defects, we also created the same type of defects on HOPG and we measured them by scanning tunneling microscopy in ambient conditions. The atomically resolved STM images shows a characteristic $\sqrt{3} \times \sqrt{3}$ pattern around the defects, similar to those induced and imaged in ultrahigh vacuum [35] for point defects.

For the experiments we started with pristine graphene drumheads, we determined its mechanical properties and subsequently increased the defect density by $Ar^+$ ion bombardment in successive small doses. In this way, we could follow the dependence of the elastic constants with increasing defect density on the same drumhead, hence minimizing the dispersion due to changes from sample to sample.

**Results**

*TEC vs defect density*

In order to measure the TEC of graphene we devised a novel method; we measured the stress on graphene drumheads at different temperatures by AFM indentations. As the edges of the drumheads were robustly clamped to the edges of the well, variations of the membrane size with the temperature translated in a change of the stress on the suspended graphene membranes (Figure 2a). As shown previously, it is possible to extract the stress of the membrane from the linear term of the fitting of the indentation

curves to equation 1. The TEC ($\alpha$) should depend on the variation of the prestress ($\Delta\sigma$) as:

$$\alpha = \frac{\Delta\sigma}{E_{2D}\Delta T} \quad (3)$$

Where $\alpha$ is the TEC, $E_{2D}$ is the 2D Young´s modulus, $\sigma$ is the prestress of the membrane (stress at zero indentation) and $T$ is the temperature. Therefore, we were able to quantitatively measure the TEC of graphene from the lineal fitting of the dependence of the initial stress of graphene drumheads with temperature (figure 2.b). The substrate also expanded with temperature, which translated on an increase of the size of the wells. To evaluate how important is the increase of size of the boundary conditions we simulated with COMSOL Multiphysics different diameter wells on 300nm SiO$_2$ on a substrate of silicon with 0.25x0.25 mm$^2$ of size and 0.5mm thick. We observed that the change in the size of the boundaries with temperature was negligible, because of the small TEC of SiO$_2$; variations of the size of boundary conditions lead to an error in the measurement of the TEC of around 1.5·10$^{-7}$K$^{-1}$ for a 1µm radius hole, which is an order of magnitude below the values we measured.

Figure 2.a shows stress *vs*. temperature data obtained from indentations on pristine and defective membranes. Green data and linear fitting correspond to the temperature dependence of the stress of a pristine membrane. The observed variation in prestress indicated that graphene contracted with temperature, corresponding to a negative thermal expansion coefficient of the membrane. Fitting the data to Equation 3 yields a TEC of -7.5·10$^{-6}$K$^{-1}$. The average value obtained for the TEC in more than 10 different pristine drumheads was (-8±1)·10$^{-6}$K$^{-1}$. This value agreed with previous values obtained for the TEC of pristine graphene by different techniques [9, 38] and validated the technique used in this work.

Upon Ar$^+$ bombardment we tracked the dependence of the TEC with the defect density. We observed that the increase of the density of defects resulted in a decrease of the slope of the stress *vs*. temperature plots. This suggested that the TEC of graphene tends to zero as we increased the density of vacancies (figure 2.c). We presented Molecular Dynamics (MD) simulations to understand the underlying mechanism that removed the rippling. Simulations showed a reduction of the thermal contraction of the membranes with the introduction of single atom vacancies, in good agreement with the experiments. In addition, simulations revealed that the subtraction of an atom from the graphene lattice caused a distortion around the defect that lead to the emergence of strain fields with a value of 0.7% in the vicinity of the vacancy. These strain so induced are high enough to significantly reduce rippling (as in a wrinkled paper). Indeed it was theoretically studied previously, that a 1% of strain completely quench anharmonic thermal fluctuations in graphene [39]. To sum up, simulations and experiments pointed towards a reduction of rippling by vacancies due to the strain fields created around them. Consequently they decreased the thermal contraction of graphene. Additional details can be found in reference [25].

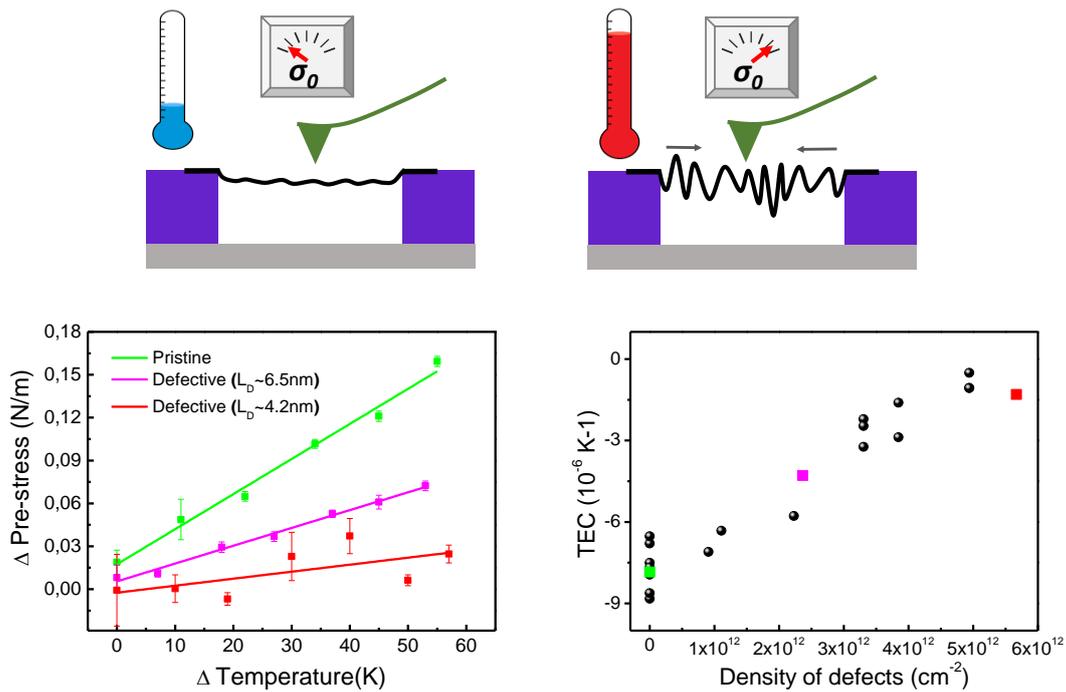

**Figure 2** *a) Scheme of the experiment performed to measure the thermal expansion coefficient of graphene. The increase of temperature on the drumhead produces an increase of the stress that we can measure by AFM indentations. b) Temperature dependence of the stress measured at different vacancy concentrations. c) Dependence of the TEC as a function of the defect density. Colored dots correspond to the data obtained from the curves in panel b.*

## *Young´s modulus vs. defect density*

The reduction of the thermal contraction coefficient with increasing vacancy density pointed towards quenching of rippling by single vacancy defects. Therefore, with the aim of quantifying the effect of rippling on the elastic properties of graphene we experimentally measured the dependence of the Young´s modulus with the defect density. In order to obtain the Young´s modulus of the membranes we performed indentation experiments on graphene drumheads and fitted the data to equation 2. Experiments performed in more than 15 pristine membranes leaded to a Young´s modulus of 300±40 N/m , in good agreement with previous experiments [1]. The initial stress of the drumheads ($\sigma_0$) was in the range of 0.05 - 0.8 N/m, typical in this kind of experimental set ups. We introduced a controlled density of single atom vacancies in the graphene membranes by Ar$^+$ ion bombardment. Figure 2.a shows two indentation curves performed on the same graphene drumhead. The green curve was carried out on the membrane as deposited (pristine), and the red curve after introducing 4x10$^{12}$ vacancies/cm$^2$. Figure 2.b shows a histogram of the Young's modulus obtained from different curves acquired on the pristine and defective membranes. As can be observed, the defective membrane was found to be stiffer (485N/m) than the pristine one (305N/m).

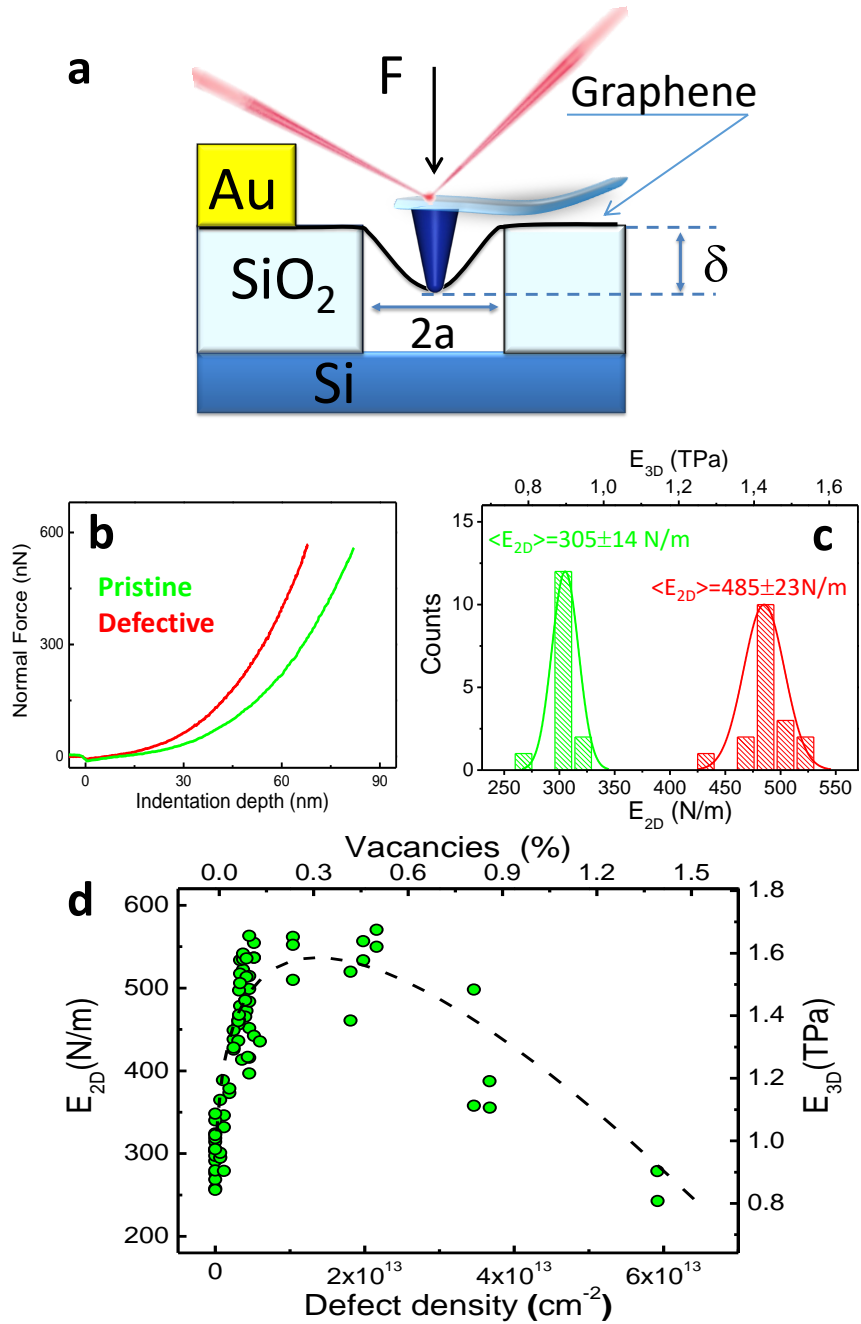

*Figure 3* a) Scheme of the experiments performed to extract the Young´s modulus of the membranes. b) Indentation curves performed on the same drumhead pristine (green) and with $4\times10^{12}$ vacancies/cm$^2$ (red). c) Histogram of the Young's modulus obtained from different curves performed on the same drumhead of panel b. d) Young's modulus as a function of the density of defects obtained from more than 10 different drumheads.

We repeated the experiments for 10 membranes finding a consistent increase of graphene stiffness at low densities of vacancies (figure 2.c). We observed an almost two-fold increase of the Young´s modulus (up to ~550 N/m) with approximately a 0.2% of single carbon vacancies. We attributed the initial increase to a reduction of rippling by defects, in good agreement with the previous experiment. On the other hand, for defect densities higher than 0.2% we observed a decrease of the stiffness as we increased the density of vacancies, which we attributed to the predicted structural damage caused by the defects. The results presented in this section are published and discussed in more detail in reference [24].

*Young´s modulus vs. strain*

As the strain fields around vacancies reduce the TEC in absolute value, global strain on graphene membranes should also remove rippling. Roldán et al. studied the effect of the strain on thermal fluctuations of graphene with a combination of atomistic Monte Carlo simulations and elastic theory of membranes within the self-consistent screening approximation [39]. They observed that strain suppressed anharmonic coupling between bending and stretching modes. Concretely a 1% of strain completely destroyed the anharmonic thermal fluctuations in graphene. In order to confirm rippling as the origin of the enhancement of the Young´s modulus of graphene with defects we also measured the dependence of the Young´s modulus of graphene by indentation on blisters. Notice that using a purely indentation scheme is very difficult to achieve high average strains across the membrane. As already discussed, point forces induce high strain below the tip apex which easily overcome the critical strength for relatively low average strain.

To apply global strain on suspended graphene membranes we used a simple variation on the previously described indentation setup. These drumheads are well-known to behave as sealed micro-chambers and be able to retain pressure for long periods of time [40]. We used an AFM in a chamber where we were capable of varying the pressure from vacuum ($10^{-6}$ mbar) to an overpressure of 3 bars and measured simultaneously the topography and mechanical properties of the graphene blisters.

The evidence for the overpressure was nicely captured in AFM topographies showing combed drumheads (figure 4.a). Then we performed indentation experiments on these bulged drumheads. The sealing of these micro-chambers was not perfect and the pressure inside tended to equal that of the surrounding atmosphere [40, 41]. Taking advantage of this, we introduced pressure on the chamber and made topographies and indentation curves as a function of time. The analysis of the indentation performed on graphene blisters is not as straightforward as the ones performed on flat membranes. Equation (1) is not an exact solution of von Karman equation. It can be rather seen as some kind of polinomial approximation that fulfils exact results for lower indentation regime in a membrane with prestress (linear term) and high indentation regime for a membrane without prestress (cubic term). A clear consequence of the membrane combing is that odd symmetry of the system is destroyed and hence some other even terms should appear (now, the geometry varies for indentations on one side to the other of the membrane as the membrane is concave or convex depending on the side). Hence, equation (1) is not valid any longer and we had to find a proper manner to extract the $E_{2D}$ from the indentation curves. In order to get an accurate approach to analyze the curves we performed finite element simulations (COMSOL Multiphysics). We started by simulating a circular clamped membrane under a hydrostatic pressure between 0 and 4 bar indented at the center. From the simulations we plot force *vs.* indentation curves analogous to the experiments. We fit the numerical curves to a full 3$^{rd}$ order polynomial and try to obtain the $E_{2D}$ from the cubic term as $C_3=E_{2D}/a^2$ at different applied forces. Note again, that as a consequence of the symmetry breaking, now this full 3$^{rd}$ order polynomial is more natural than just a polynomial with odd terms. From this analysis we can plot the 2D Young's modulus obtained from $C_3$ as a function of the maximum force applied in the indentation. It is necessary to first specify variables and properties of the elements in the simulation, so the Young's modulus of the membrane is a known input value of the simulation. This allowed us to compare the Young´s modulus set as an input with the value obtained from the analysis of the curve extracted from the simulation. At very low forces the Young´s modulus obtained was much higher than the one introduced as an input. This is because the strain induced by pressure at the same deflection is much higher than the strain created by the indentation. However, the value obtained of $E_{2D}$ rapidly tends to the input value at moderate forces (200-500nN), easily achievable in our experiments. Therefore, we used the same analysis for the experimental curves.

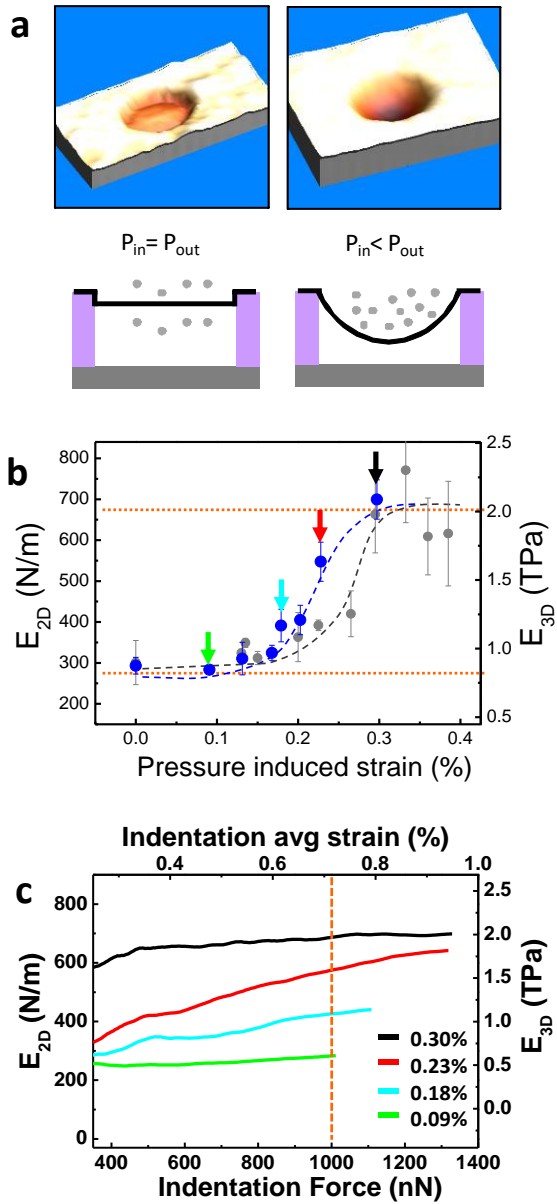

*Figure 4 a) AFM image and scheme of a blister with both sides at the same pressure (left), a blister with higher outer pressure (right). b) $E_{2D}$ vs pressure induce strain obtained from 2 different drumheads. Dashed red lines are to guide the eye. c) $E_{2D}$ obtained as a function of the applied force during indentation from 4 curves obtained at different pressures. Colors correspond to the colored arrows in panel b. Orange dashed line corresponds to 1000nN, the maximum force selected for fitting the data.*

Figure 4.b shows the dependence of the $E_{2D}$ with strain induced by pressure (total strain is the sum of strain caused by pressure plus initial membrane strain). At 0% of pressure induced strain we measured the typical value for the Young's modulus of pristine graphene, which is about 300N/m. For low strains (0-0.15%) we measured the same value with no variations. This is consistent with the measurements reported by several groups showing that Young's modulus is independent of the initial strain of the flake. When the strain induced by pressure is around 0.15% we started to infer an increase in the stiffness of the membrane. Above 0.3% we observed a third region where the $E_{2D}$

remains constant at ~680N/m. It is also important to take into account that to measure the Young´s modulus of the membrane, during the indentation, graphene is also strained. Figure 4.c illustrate the dependence of the $E_{2D}$ obtained from 4 representative experimental curves as a function of the maximum force taken for the analysis, thus the same analysis performed with the curves obtained by COMSOL Multiphysics. All the curves showed, at forces below 400nN (hidden in the figure) the same maximum of $E_{2D}$ vs. maximum force than in the analysis of the simulated curves. However, in contrast with the simulations, we observed a different behavior; the curves obtained at different pressures did not converge to the same value at high forces, but had different trends depending on the strain of the bulged drumhead. When the initial strain of the drumhead was low (<0.15%) the $E_{2D}$ obtained remained constant at ~300N/m for high forces. For intermediate strains the obtained $E_{2D}$ increased with the applied force; for high strains it was again flat but at a value approximately twice the initial value (~700N/m).Therefore, the tendency of the $E_{2D}$ vs indentation force was consistent with the behavior observed at different strains. We always fitted our curves at a maximum force of 1000nN as a criterion to analyze the data. See reference [26] for a more extended discussion on this experiment.

**Discussion**

The first experiment here presented showed a reduction of the thermal contraction of graphene with the vacancy content of almost a 100% at a defect density of ~$5x10^{12}$def/cm$^2$. In [25] we attributed this decrease to the quenching of thermal rippling by defects. Molecular dynamics simulations showed the same tendency and pointed towards relatively high strains (0.7%) created around defects as the cause of rippling reduction. This strain was high enough to significantly diminish all the long wavelength rippling (that have the most important influence on the negative TEC of graphene), localizing ripples between vacancies. On the other hand, other defect as sp$^3$ defects or inversed Stone-Wales defects, create some out of plane corrugation [42], which contributes to soften the graphene membranes. Thus, the nature of the defects is critical, because different type of defect would lead to opposite results on the mechanics of graphene. As described in methods, and in more detail in reference [24], Raman spectroscopy an STM images indicate creation of single atom point defects with sp$^2$ hybridization. In conclusion, the single-vacancy nature of defects was determined to be very important to reduce rippling, as other type of defects may not create such a high strain in their surroundings. Notice that a similar effect could be also caused to static rippling.

The second experiment [24] showed the dependence of the $E_{2D}$ with the vacancy content. We observed an initial increase of the $E_{2D}$ for low densities of defects up to ~$5x10^{12}$ def/cm$^2$ where it reaches ~550N/m, almost two times higher than that measured in pristine membranes. As rippling reduced the $E_{2D}$ of membranes, quenching of ripples by defects would result in an increase of the $E_{2D}$. On the other hand, for higher vacancy content we observed a decrease of the $E_{2D}$ with the defect density. Contrary to the initial stiffening, this decrease was predicted by all previous theoretical works showing the dependence of the Young's modulus with the density of single vacancies by atomistic simulations, both first principles and molecular dynamics [43-45]. The apparent disagreement at low densities of defects is because in these studies only the contribution of the defects to the structural stiffness of the membranes was taken into account, but not the possible thermodynamic effects (ripples) on the mechanical properties of graphene. Precise determination of the effects of rippling on the elastic constants of graphene would require dynamic atomistic simulations of micron-sized membranes, which is not feasible nowadays or, as we will see bellow, reproduction of the intrinsic rippling observed in free-standing graphene [17].

These two mechanisms affecting the mechanical properties of graphene could be captured by a qualitative model, which gives the equation:

$$E_{2D} = K \left(\frac{1}{l_0} + n_i\right)^{\eta/2} \left(1 - c\left(\frac{1}{l_0} + n_i\right)\right) \quad (4)$$

where $K$ and $c$ are constants, $l_0$ is the localization length for flexural phonons in pristine graphene and $n_i$ is the density of defects induced by irradiation. Therefore the first term accounts for the stiffening caused by reductions of rippling, while the second is a consequence of the softening due to the removal of carbon-carbon bonds. The dashed line in Fig. 3d depicts a fitting to our experimental results according to equation (4). Best fitting to our experimental data yields K =1.5×10$^9$ N m$^{\eta-2}$, l$_0$ =50 nm, η =0.36 and c =1.2×10$^{-18}$ m$^2$

By last, we provided the dependence of the $E_{2D}$ with an induced global strain in our graphene membranes[26]. Here we observed an increase and later saturation of the $E_{2D}$. This observation matched with the scenario of mechanical properties strongly influenced by rippling. As ripples soften membranes, the removal of rippling by the effect of high global strain induced by pressure results in an effectively stiffening of graphene. The increase of $E_{2D}$ with the indentation force at intermediate strains is also consistent with the picture of mechanical properties influenced by rippling, as the membrane is also strained during the indentation. On the other hand, the initial plateau where the $E_{2D}$ remains constant indicates that there is an infrared cut-off of the rippling, which means that the ripples with longer wavelength of a certain value do not exist in our suspended graphene membranes, or at least their influence to the $E_{2D}$ is negligible. As shown in reference [39] low strain start quenching anharmonicities of long wavelength, suppressing lower wavelength ripples when increasing the strain.

Figure 5 summarizes the three results described in this manuscript where we can clearly observe consistent numerical values, i.e. the maximum of the $E_{2D}$ *vs.* defect density (red) matches with the density of vacancies where the absolute value of the TEC reaches the minimum (green) (~5x10$^{12}$ defects/cm$^2$). Moreover, very low defect content (<10$^{12}$ defects/cm$^2$) has undetectable effects on either of the two magnitudes. In addition, the increase of the Young's modulus achieved introducing defects is a ~15% smaller but with similar tendency to the observed with strain. The higher Young's modulus reached with strain makes sense as single vacancies also contribute to structurally reduce the stiffness of graphene and might not cancel rippling fully. These coincidences strongly suggest that all the phenomena described here rely on the same physics.

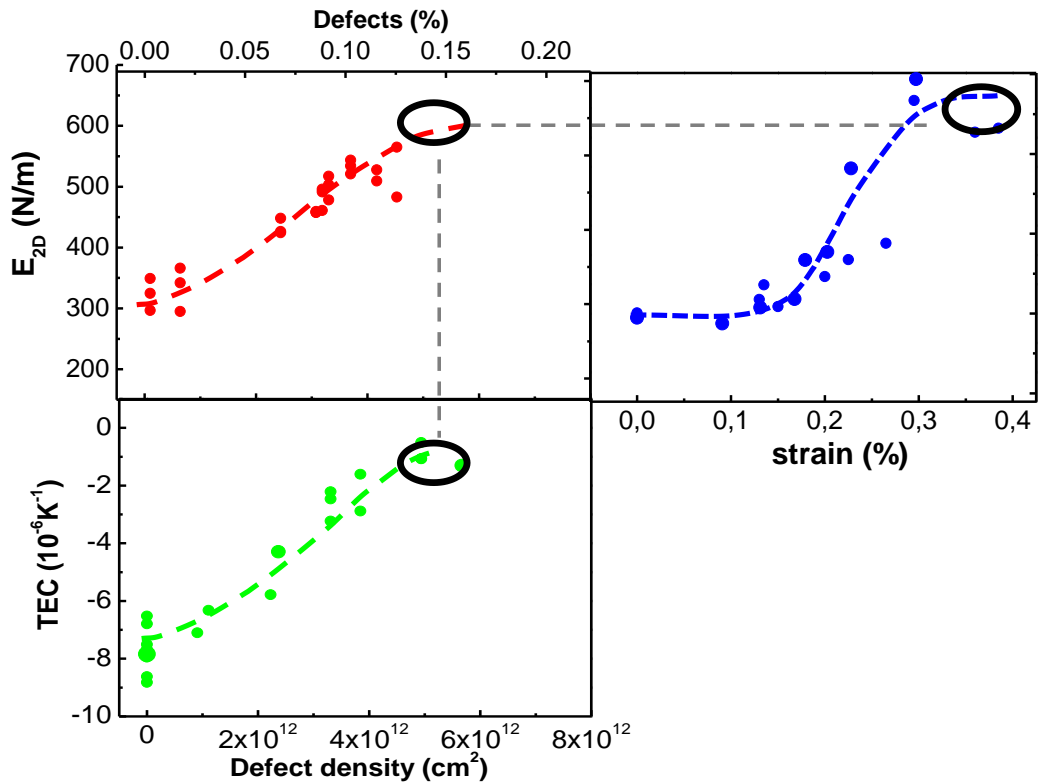

*Figure 5* In red (top left panel) $E_{2D}$ vs. defect density, in blue (top right panel) $E_{2D}$ vs. strain and in green (bottom panel) TEC vs. density of defects. The scales of the x and y axes coincide with that of the adjacent panels to show that the values coincide. Dashed lines are to guide the eye to show that the defect density of the maximum $E_{2D}$ and the minimum absolute value of the TEC coincide, and the maximum of the $E_{2D}$ with strain and defects it is also similar.

At first sight, all these behaviors can be qualitatively explained as the reduction of out of plane thermal fluctuations by strain and defects. However, some subtle experimental observations are discordant with the scenario of just thermal fluctuations. If we plot the Young´s modulus and the TEC as a function of the distance between defects (figure 6) it is clear that, when the distance between defects is higher than 10 nm, we obtained the same values as in pristine graphene for both magnitudes. Note that pristine membranes, without defects, are represented in the plot with a mean distance between defects of 50nm (the maximum distance that is possible for Raman to detect), but the real distance for microexfoliated flakes, which always exhibit a high degree of crystallinity/perfection, is expected to be much higher. TEC dependence with the mean distance between defects also indicates that defects need to be closer than 10 nm to have perceptible effects on the rippling. There are two possible causes that can give this pronounced decay with the density of defects. Firstly, that defects fully localize the rippling between vacancies and as a consequence the maximum wavelength of the rippling existing in graphene is around 10nm. Secondly, that several vacancies are required to remove the long-wavelength rippling. Accordingly, low densities of defects would produce subtle variations of the $E_{2D}$. Indeed, the mechanism behind the suppression of rippling by the local strain around vacancies is not well studied and new theoretical development on this topic would be relevant.

In contrast, the way global strain removes rippling is more direct and well known. The initial plateau observed in the $E_{2D}$ as a function of the strain points also towards to a cut-off in the thermal fluctuations of about 10 nanometers, as pointed by self-consistent approximation [39, 46]. On the other hand, it is also important to note that thermal fluctuations with short wavelength have lower consequences on the elasticity and thermo-mechanics of membranes [39, 47]. In contrast, we show that the influence of rippling on the mechanics of graphene is extremely high: the thermal contraction is as high as $-8 \times 10^{-6} K^{-1}$ and we measured a reduction of almost a 100% and a similar relative increase for the Young modulus, contrary to what is expected from thermal fluctuations with wavelength <10nm. As an example, MD simulations always predict a lower absolute value of TEC than the experimental value. This is usually justified by the small size of the simulated membranes compared to the experiments, which defines a cut off in the rippling. However, the MD simulations we presented [25] were performed on a membrane of 25nm x 25nm. This size doubles the cut off that explains our experimental results (~10nm), but simulations give a TEC of around $-3 \times 10^{-6} K^{-1}$, which is less than a half of the experimental value. On the other hand, anharmonicity introduces a wavevector-dependent elastic modulus, $E_{2D} \propto q^{\eta}$ where $\eta \sim 0.36$ being q is the momentum or inverse length of flexural phonons. Self-consistent approximation calculations predicted a renormalization of the $E_{2D}$ lower than a 5% caused by the oscillations of this wavelength at room temperature. Moreover, if long wavelength fluctuations would have such large effects as predicted by theory and measured in the experiments presented here, their effect on elasticity of graphene membranes would have been observed in simple experiments, such as elastic modulus dependence on the size of micron-scaled samples, or a strong dependence of elasticity at low strains (<0.2%). In contrast, we did not observe any tendency of $E_{2D}$ vs. drumhead size [24]. Also, an increase of the Young´s modulus as a function of the initial stress or the maximum indentation depth should be also measureable (prestress in graphene drumheads varies typically between 0.05 and 0.8N/m).

In summary, out of plane thermal fluctuations could address some of the observed tendencies qualitatively, but our experiments show that only rippling with wavelength of 5-10nm affect the $E_{2D}$ or the TEC. Simultaneously theory predicts imperceptible effects from fluctuations of these small wavelengths. On the contrary, experiments show a significant negative TEC and a renormalization of almost a factor of two of the in-plane Young's modulus and we don't observe any low-stress or size dependence of either of the two magnitudes.

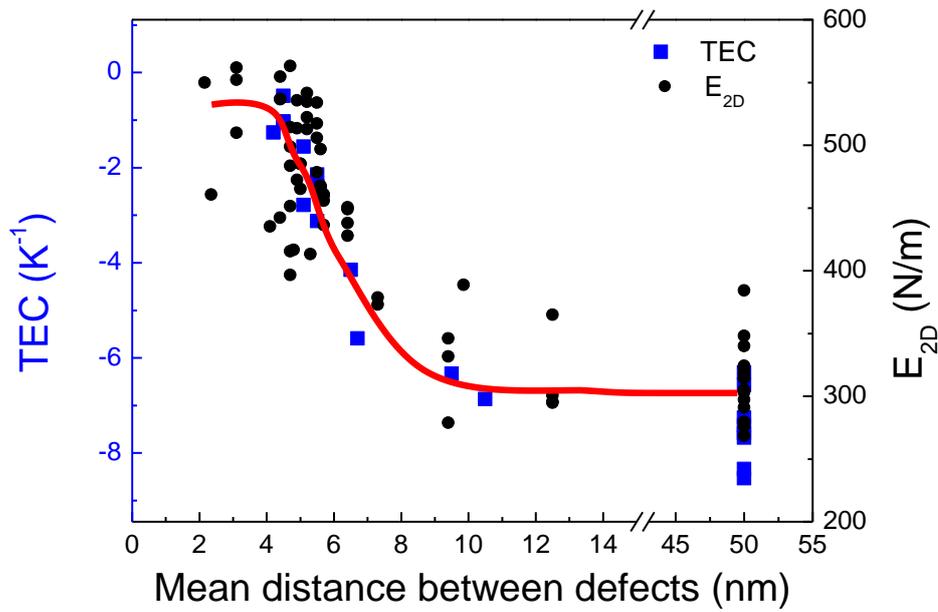

**Figure 6** *TEC (blue squared dots) and $E_{2D}$ (black circled dots) as a function of the mean distance between defects. Pristine membranes are represented as a mean distance between defects of 50 nm. It can be observed that at a distance between defects of around 10nm none of the two magnitudes show a clear difference. On the other hand, for distances between 5 and 10 nm we observe a variation of almost a factor of 2 in the $E_{2D}$ and a reduction of almost a 100% in the thermal contraction.*

To account for all the facts described above, here we propose static intrinsic rippling instead of dynamic thermal fluctuations as the governing mechanism for the renormalization of the elastic constants and the reduction of the negative TEC of graphene. Experimental [28] and theoretical works [20] suggest that static rippling may be the main cause of all the results presented here. In 2007, it was published that based on Transmission Electron Microscopy data suspended graphene exhibited static intrinsic out-of-plane deformations of 5-10nm wavelength and up to 1nm height [17, 28, 48, 49]. Ripples of this wavelength could also be observed in supported graphene [50, 51]. Among other theoretical explanations, F.L. Braghin and N. Hasselmann proposed that these atomic scale ripples observed in suspended graphene could be naturally explained as just the real-space manifestation of the Ginzburg scale [20, 21, 27]. Ginzburg length gives the size of the minimum membrane size required to have interaction between in plane and out of plane phonons, i.e. anharmonic fluctuations. These phonons in graphene are highly energetic due to its high in-plane stiffness. As explained in the introduction static ripples increase the bending rigidity of a membrane, and consequently increase the energy of thermal fluctuations with higher wavelength than the periodicity of the ripples. The work by Hasselmann et al. [20] suggests that membranes tend to crumple at this wavelength and prevent dynamic thermal fluctuations with longer wavelength to exist in graphene, which are energy costly anharmonic oscillations. The existence of this intrinsic ripples could account, even better than thermal fluctuations, our experimental observations on the behavior of the $E_{2D}$ with induced strain and defects: Firstly, mean distances between defects higher than the typical wavelength of the ripples (5-10nm) will not have relevant effects on the mechanics of graphene membranes. This accounts for the fact that we did not observe any change in the Young's modulus or the TEC with defect distances higher than 10 nm. Also it would explain the initial plateau of $E_{2D}$ observed for low strains, which is also a signature of a cut-off in the wavelength of the ripples of ~10nm, as obtained from self-consistent approximation calculations [39, 46]. Secondly, these calculations predict variations of the $E_{2D}$ of <5% from thermally activated flexural phonons with wavelength below 10nm [39, 46]. In contrast, we measured differences in the Young´s modulus approximately of a factor of 2. The amplitude of static ripples has been measured to be as high as ~1nm, which could address such a significant influence on the mechanical properties of graphene.

At first sight one might think that static rippling could not account for the observed changes in the TEC of defective graphene. It is well accepted that the negative TEC of graphene is due to out of plane thermal fluctuations. But, as mentioned above, static ripples renormalize all the elastic constant of membranes and in particular the bending rigidity of graphene. Consequently fluctuations with higher wavelength than the ripples would be lessened (even suppressed) and makes impossible to justify such a significant negative TEC from dynamic fluctuations. However, static rippling of graphene membranes at the Ginzburg length would imply a dependence of the wavelength of the ripples with the temperature; Ginzburg length decreases with temperature as $L_G \propto \kappa/\sqrt{E_{2D} k_B T}$ [20]. Hence, this would make graphene membranes to contract with the temperature in order to allow shorter wavelength ripples, resulting in a negative thermal coefficient. In fact D.A. Kirilenko et al. reported in 2011 the dependence of the rippling with temperature measured by TEM diffraction experiments [28]. Surprisingly, they observed a decrease of the amplitude of rippling with increasing temperature. However, their results were compatible with membrane contraction with temperature due to a decrease of the wavelength of rippling with temperature. This observation was unexpected but it is in good agreement with our interpretation based in the experimental data here presented and supports that TEC of graphene is negative due to formation of ripples with a wavelength equal to the Ginzburg length, instead of the well accepted dynamic flexural phonons. It is remarkable that there is still no experimental evidence of the size or stress dependence of the negative TEC of graphene.

Other previous works explored the influence of wrinkling on the mechanical properties of graphene. They showed a significant decrease in the in-plane Young's modulus [14, 16] and an important increase of the bending rigidity [15]. A softening of graphene membranes with temperature was also reported [52]. However, in these studies graphene membranes were obtained by CVD deposition and later transferred to a substrate with wells. Consequently, they exhibited micron-sized wrinkles and crumples due to the sample preparation procedure and presented one order of magnitude lower $E_{2D}$ than that accepted for plane graphene. We have to emphasize the well differentiated effects observed in such samples from that observed in our suspended graphene that was prepared directly by mechanical exfoliation on circular wells and showed initially the well-accepted 350N/m of $E_{2D}$ for 'flat' pristine graphene. Therefore our results are akin to be related to intrinsic mechanisms rather than external crumpling.

What still remains difficult to explain is the high value of the non-renormalized or 'bare' $E_{2D}$ of graphene; the experiments shown in this manuscript point towards an $E_{2D}$ of ~650N/m, almost two times higher than the main value obtained by DFT that is ~350N/m. Nevertheless, theoretical works present high dispersion and the maximum theoretical value obtained from first principles is ~420N/m [53]. A non-discardable reason for this apparent disagreement could be related to the factor q(v) used in eq. 1 for the determination of $E_{2D}$. The factor q(v) is an approximation to make this valid for every v, but as shown in reference [26] finite element simulations gives a factor slightly different. Out of plane structures, as discuss here, should also affect the value of the Poisson ratio value even to negative values. Additionally, in-depth theoretical studies about indentation experiments show that the geometry and size of the tip could affect significantly to the equation governing the experiment [31]. Therefore, in this manuscript we do not aim to give a precise result of the 'bare' $E_{2D}$ of graphene but to provide a qualitatively comparison of the elastic properties of graphene when defects or strain are induced.

**Conclusions**

To sum up, here in we presented three experimental results related with the membrane character of graphene. We measured a reduction of almost a 100% of the thermal contraction of graphene with the introduction of single atomic vacancies. We also observed an enhancement of a factor of two on the elastic response of graphene with the induction of atomic defects and strain. At a mean distance between defects of 4-5nm we observed the maximum Young´s modulus and maximum value of TEC reaching almost zero. Additionally, the maximum value for $E_{2D}$ obtained by introducing defects in the graphene lattice is also similar to that obtained for our highest global induced strains. All these observations could be explained by the suppression of out of plane structures in graphene by defects and strain (ripples, wrinkles, perpendicular phonons…). Molecular dynamics simulations revealed that the suppression of out of plane structures by defects is due to strain fields created around the single vacancies. Our data showed no variation for mean distances between defects higher than 10nm and strains below 0.3%. This points towards an infrared cut-off in the out of plane structures of around 10nm. Dynamic thermal fluctuations with shorter wavelength than 10nm should have undetectable effects in both $E_{2D}$ and TEC. Based on recent theoretical works we propose static rippling observed in suspended graphene as the governing mechanism, not only for the renormalization of the elastic constants, but also for the negative TEC of graphene. This explanation of our results has a profound impact on the understanding of mechanics and thermomechanics of graphene.

*Acknowledgements*

We acknowledge financial support from the Spanish MINECO through PID2019-106268GB-C31 and from Comunidad de Madrid (S2018/NMT-4511, NMAT2D-CM)

JG-H and CG-N acknowledge support from the Spanish Ministry of Science and Innovation, through the "María de Maeztu" Programme for Units of Excellence in R&D (CEX2018-000805-M). GLP acknowledge support from the Spanish Ministry of Science and Innovation for the JdC Fellowship FJCI-2017-32370.

**References**

1. Lee, C.; Wei, X.; Kysar, J. W.; Hone, J., Measurement of the elastic properties and intrinsic strength of monolayer graphene. *science* **2008,** *321* (5887), 385-388.
2. Lindahl, N.; Midtvedt, D.; Svensson, J.; Nerushev, O. A.; Lindvall, N.; Isacsson, A.; Campbell, E. E., Determination of the bending rigidity of graphene via electrostatic actuation of buckled membranes. *Nano letters* **2012,** *12* (7), 3526-3531.
3. Mermin, N. D., Crystalline order in two dimensions. *Physical Review* **1968,** *176* (1), 250.
4. Katsnelson, M. I., Graphene: carbon in two dimensions. *Materials today* **2007,** *10* (1-2), 20-27.
5. Nelson, D. R.; Piran, T.; Weinberg, S., *Statistical mechanics of membranes and surfaces*. World Scientific: 2004.
6. Nelson, D.; Peliti, L., Fluctuations in membranes with crystalline and hexatic order. *Journal de physique* **1987,** *48* (7), 1085-1092.
7. Helfrich, W., Elastic properties of lipid bilayers: theory and possible experiments. *Zeitschrift für Naturforschung C* **1973,** *28* (11-12), 693-703.
8. Peliti, L.; Leibler, S., Effects of thermal fluctuations on systems with small surface tension. *Physical review letters* **1985,** *54* (15), 1690.
9. Bao, W.; Miao, F.; Chen, Z.; Zhang, H.; Jang, W.; Dames, C.; Lau, C. N., Controlled ripple texturing of suspended graphene and ultrathin graphite membranes. *Nature nanotechnology* **2009,** *4* (9), 562-566.
10. Cai, Q.; Scullion, D.; Gan, W.; Falin, A.; Zhang, S.; Watanabe, K.; Taniguchi, T.; Chen, Y.; Santos, E. J.; Li, L. H., High thermal conductivity of high-quality monolayer boron nitride and its thermal expansion. *Science advances* **2019,** *5* (6), eaav0129.
11. De Andres, P.; Guinea, F.; Katsnelson, M., Bending modes, anharmonic effects, and thermal expansion coefficient in single-layer and multilayer graphene. *Physical Review B* **2012,** *86* (14), 144103.
12. Jiang, J.-W.; Wang, J.-S.; Li, B., Thermal expansion in single-walled carbon nanotubes and graphene: Nonequilibrium Green's function approach. *Physical Review B* **2009,** *80* (20), 205429.
13. Gao, W.; Huang, R., Thermomechanics of monolayer graphene: Rippling, thermal expansion and elasticity. *Journal of the Mechanics and Physics of Solids* **2014,** *66*, 42-58.
14. Nicholl, R. J.; Lavrik, N. V.; Vlassiouk, I.; Srijanto, B. R.; Bolotin, K. I., Hidden area and mechanical nonlinearities in freestanding graphene. *Physical review letters* **2017,** *118* (26), 266101.
15. Blees, M. K.; Barnard, A. W.; Rose, P. A.; Roberts, S. P.; McGill, K. L.; Huang, P. Y.; Ruyack, A. R.; Kevek, J. W.; Kobrin, B.; Muller, D. A., Graphene kirigami. *Nature* **2015,** *524* (7564), 204-207.

16. Nicholl, R. J.; Conley, H. J.; Lavrik, N. V.; Vlassiouk, I.; Puzyrev, Y. S.; Sreenivas, V. P.; Pantelides, S. T.; Bolotin, K. I., The effect of intrinsic crumpling on the mechanics of free-standing graphene. *Nature communications* **2015,** *6* (1), 1-7.
17. Meyer, J. C.; Geim, A. K.; Katsnelson, M. I.; Novoselov, K. S.; Booth, T. J.; Roth, S., The structure of suspended graphene sheets. *Nature* **2007,** *446* (7131), 60-63.
18. Fasolino, A.; Los, J.; Katsnelson, M. I., Intrinsic ripples in graphene. *Nature materials* **2007,** *6* (11), 858-861.
19. Hu, Y.; Chen, J.; Wang, B., On the intrinsic ripples and negative thermal expansion of graphene. *Carbon* **2015,** *95*, 239-249.
20. Braghin, F.; Hasselmann, N., Thermal fluctuations of free-standing graphene. *Physical Review B* **2010,** *82* (3), 035407.
21. Gornyi, I.; Kachorovskii, V. Y.; Mirlin, A., Rippling and crumpling in disordered free-standing graphene. *Physical Review B* **2015,** *92* (15), 155428.
22. Gazit, D., Correlation between charge inhomogeneities and structure in graphene and other electronic crystalline membranes. *Physical Review B* **2009,** *80* (16), 161406.
23. San-Jose, P.; González, J.; Guinea, F., Electron-induced rippling in graphene. *Physical review letters* **2011,** *106* (4), 045502.
24. López-Polín, G.; Gómez-Navarro, C.; Parente, V.; Guinea, F.; Katsnelson, M. I.; Perez-Murano, F.; Gómez-Herrero, J., Increasing the elastic modulus of graphene by controlled defect creation. *Nature Physics* **2015,** *11* (1), 26-31.
25. López-Polín, G.; Ortega, M.; Vilhena, J.; Alda, I.; Gomez-Herrero, J.; Serena, P. A.; Gomez-Navarro, C.; Pérez, R., Tailoring the thermal expansion of graphene via controlled defect creation. *Carbon* **2017,** *116*, 670-677.
26. López-Polín, G.; Jaafar, M.; Guinea, F.; Roldán, R.; Gómez-Navarro, C.; Gómez-Herrero, J., The influence of strain on the elastic constants of graphene. *Carbon* **2017,** *124*, 42-48.
27. Hasselmann, N.; Braghin, F., Nonlocal effective-average-action approach to crystalline phantom membranes. *Physical Review E* **2011,** *83* (3), 031137.
28. Kirilenko, D.; Dideykin, A.; Van Tendeloo, G., Measuring the corrugation amplitude of suspended and supported graphene. *Physical Review B* **2011,** *84* (23), 235417.
29. Ferrari, A. C.; Meyer, J.; Scardaci, V.; Casiraghi, C.; Lazzeri, M.; Mauri, F.; Piscanec, S.; Jiang, D.; Novoselov, K.; Roth, S., Raman spectrum of graphene and graphene layers. *Physical review letters* **2006,** *97* (18), 187401.
30. Komaragiri, U.; Begley, M.; Simmonds, J., The mechanical response of freestanding circular elastic films under point and pressure loads. *J. Appl. Mech.* **2005,** *72* (2), 203-212.
31. Vella, D.; Davidovitch, B., Indentation metrology of clamped, ultra-thin elastic sheets. *Soft Matter* **2017,** *13* (11), 2264-2278.
32. Wei, X.; Fragneaud, B.; Marianetti, C. A.; Kysar, J. W., Nonlinear elastic behavior of graphene: Ab initio calculations to continuum description. *Physical Review B* **2009,** *80* (20), 205407.
33. Lin, Q.-Y.; Jing, G.; Zhou, Y.-B.; Wang, Y.-F.; Meng, J.; Bie, Y.-Q.; Yu, D.-P.; Liao, Z.-M., Stretch-induced stiffness enhancement of graphene grown by chemical vapor deposition. *Acs Nano* **2013,** *7* (2), 1171-1177.
34. Gómez-Navarro, C.; De Pablo, P. J.; Gómez-Herrero, J.; Biel, B.; Garcia-Vidal, F.; Rubio, A.; Flores, F., Tuning the conductance of single-walled carbon nanotubes by ion irradiation in the Anderson localization regime. *Nature materials* **2005,** *4* (7), 534-539.
35. Ugeda, M. M.; Brihuega, I.; Guinea, F.; Gómez-Rodríguez, J. M., Missing atom as a source of carbon magnetism. *Physical Review Letters* **2010,** *104* (9), 096804.
36. Lehtinen, O.; Kotakoski, J.; Krasheninnikov, A.; Tolvanen, A.; Nordlund, K.; Keinonen, J., Effects of ion bombardment on a two-dimensional target: Atomistic simulations of graphene irradiation. *Physical review B* **2010,** *81* (15), 153401.

37. Eckmann, A.; Felten, A.; Mishchenko, A.; Britnell, L.; Krupke, R.; Novoselov, K. S.; Casiraghi, C., Probing the nature of defects in graphene by Raman spectroscopy. *Nano letters* **2012,** *12* (8), 3925-3930.
38. Yoon, D.; Son, Y.-W.; Cheong, H., Negative thermal expansion coefficient of graphene measured by Raman spectroscopy. *Nano letters* **2011,** *11* (8), 3227-3231.
39. Roldán, R.; Fasolino, A.; Zakharchenko, K. V.; Katsnelson, M. I., Suppression of anharmonicities in crystalline membranes by external strain. *Physical Review B* **2011,** *83* (17), 174104.
40. Bunch, J. S.; Verbridge, S. S.; Alden, J. S.; Van Der Zande, A. M.; Parpia, J. M.; Craighead, H. G.; McEuen, P. L., Impermeable atomic membranes from graphene sheets. *Nano letters* **2008,** *8* (8), 2458-2462.
41. Manzanares-Negro, Y.; Ares, P.; Jaafar, M.; López-Polín, G.; Gómez-Navarro, C.; Gómez-Herrero, J., Improved graphene blisters by ultrahigh pressure sealing. *ACS Applied Materials & Interfaces* **2020,** *12* (33), 37750-37756.
42. Banhart, F.; Kotakoski, J.; Krasheninnikov, A. V., Structural defects in graphene. *ACS nano* **2011,** *5* (1), 26-41.
43. Ansari, R.; Ajori, S.; Motevalli, B., Mechanical properties of defective single-layered graphene sheets via molecular dynamics simulation. *Superlattices and Microstructures* **2012,** *51* (2), 274-289.
44. Fedorov, A.; Popov, Z.; Fedorov, D.; Eliseeva, N.; Serjantova, M.; Kuzubov, A., DFT investigation of the influence of ordered vacancies on elastic and magnetic properties of graphene and graphene-like SiC and BN structures. *physica status solidi (b)* **2012,** *249* (12), 2549-2552.
45. Jing, N.; Xue, Q.; Ling, C.; Shan, M.; Zhang, T.; Zhou, X.; Jiao, Z., Effect of defects on Young's modulus of graphene sheets: a molecular dynamics simulation. *Rsc Advances* **2012,** *2* (24), 9124-9129.
46. López-Polín, G.; Jaafar, M.; Guinea, F.; Roldán, R.; Gómez-Navarro, C.; Gómez-Herrero, J., Strain dependent elastic modulus of graphene. *arXiv preprint arXiv:1504.05521* **2015**.
47. Zakharchenko, K.; Roldán, R.; Fasolino, A.; Katsnelson, M., Self-consistent screening approximation for flexible membranes: Application to graphene. *Physical Review B* **2010,** *82* (12), 125435.
48. Wang, W. L.; Bhandari, S.; Yi, W.; Bell, D. C.; Westervelt, R.; Kaxiras, E., Direct imaging of atomic-scale ripples in few-layer graphene. *Nano letters* **2012,** *12* (5), 2278-2282.
49. Ludacka, U.; Monazam, M.; Rentenberger, C.; Friedrich, M.; Stefanelli, U.; Meyer, J.; Kotakoski, J., In situ control of graphene ripples and strain in the electron microscope. *npj 2D Materials and Applications* **2018,** *2* (1), 1-6.
50. Haghighian, N.; Convertino, D.; Miseikis, V.; Bisio, F.; Morgante, A.; Coletti, C.; Canepa, M.; Cavalleri, O., Rippling of graphitic surfaces: a comparison between few-layer graphene and HOPG. *Physical Chemistry Chemical Physics* **2018,** *20* (19), 13322-13330.
51. Geringer, V.; Liebmann, M.; Echtermeyer, T.; Runte, S.; Schmidt, M.; Rückamp, R.; Lemme, M. C.; Morgenstern, M., Intrinsic and extrinsic corrugation of monolayer graphene deposited on SiO 2. *Physical review letters* **2009,** *102* (7), 076102.
52. Storch, I. R.; De Alba, R.; Adiga, V. P.; Abhilash, T.; Barton, R. A.; Craighead, H. G.; Parpia, J. M.; McEuen, P. L., Young's modulus and thermal expansion of tensioned graphene membranes. *Physical Review B* **2018,** *98* (8), 085408.
53. Konstantinova, E.; Dantas, S. O.; Barone, P. M., Electronic and elastic properties of two-dimensional carbon planes. *Physical Review B* **2006,** *74* (3), 035417.